\documentclass[%
 reprint,
 prl,
 amsmath,amssymb,
 aps,
]{revtex4-2}

\usepackage{graphicx}
\usepackage{dcolumn}
\usepackage{xcolor}
\usepackage{dashrule}
\usepackage{tikz}
\usepackage{bm}
\usepackage{float}
\definecolor{mygreen}{HTML}{1b9e77}
\definecolor{myorange}{HTML}{1b9e77}
\definecolor{mypurple}{HTML}{1b9e77}

\usepackage{microtype}
\usepackage{braket}
\usepackage[colorlinks=true, linkcolor=blue, citecolor=blue, urlcolor=blue]{hyperref}

\begin{document}

\preprint{APS/123-QED}

\title{Systems with Quantum Dimensions}

\author{Mikołaj Myszkowski$^{1}$\thanks{Corresponding author: \texttt{first.author@inst.edu}}, 
        Mattia Damia Paciarini$^{1}$, 
        and Francesco Sannino$^{1,2,3,4}$}

\affiliation{$^{1}$Quantum Theory Center (${\hbar}$QTC) \& D-IAS, IMADA at Southern Denmark Univ., Campusvej 55, 5230 Odense M, Denmark}
\affiliation{$^{2}$Scuola Superiore Meridionale, Largo S. Marcellino, 10, 80138 Napoli NA, Italy}
\affiliation{$^{3}$Dept. of Physics E. Pancini, Università di Napoli Federico II, via Cintia, 80126 Napoli, Italy}
\affiliation{$^{4}$INFN sezione di Napoli, via Cintia, 80126 Napoli, Italy}

\date{\today}

\begin{abstract}
We propose quantum-mechanical systems in which the number of spatial dimensions is promoted to a dynamical quantum variable, making the effective dimension state-dependent. Interestingly, systems of this form can exhibit enhanced symmetries compared to their fixed-dimensional counterparts. As an explicit example, we analyze a two-state system for which the number of spatial dimensions is represented by a quantum operator. By evaluating the corresponding partition function, we uncover a temperature-dependent effective dimension. Our framework opens a new avenue for constructing physical systems, from gravity to condensed matter, where the very notion of dimensionality becomes quantum.
\end{abstract}


\maketitle
\emph{Introduction---}The reason why our universe manifests precisely three spatial and one temporal dimension is not yet fully understood. Although numerous approaches to quantum gravity and high-energy physics suggest that the effective number of spacetime dimensions may vary with the energy scale \cite{Carlip:2017eud}, there is still no unified understanding of how dimensionality itself could emerge as a physical, dynamical quantity.

In this Letter, we propose a new framework in which the number of spatial dimensions is promoted to a quantum variable subject to the laws of quantum mechanics. This leads to what we call quantum-dimension (QD) systems. The corresponding Hilbert space is built as a direct sum of Hilbert spaces with fixed dimensionality, allowing quantum superpositions of states residing in different numbers of dimensions. Within this setup, dimensionality becomes an observable represented by a Hermitian operator, and its measured value depends on the physical state of the system.

We further show that QD systems support enhanced symmetry structures involving transformations that mix states of different dimensionality. These novel symmetries are a hallmark of the framework, since they are absent in conventional quantum systems. We demonstrate the idea by constructing and solving a two-state QD system, a generalized version of the standard qubit, whose dimensionality is quantized. The effective dimension of the system is found to vary with the energy scale.

A similar phenomenon of dimensional reduction has also been observed in various approaches to quantum gravity, such as string theory \cite{Atick:1988si,Coudarchet:2023mfs}, loop quantum gravity \cite{Modesto:2008jz,Calcagni:2014cza}, causal dynamical triangulations \cite{Ambjorn:2005db,Ambjorn:2005qt}, asymptotic safety \cite{Niedermaier:2006ns,Litim:2006dx}, noncommutative geometry \cite{Nozari:2015iba,Amelino-Camelia:2016sru}, Hořava-Lifshitz gravity \cite{Horava:2009if,Horava:2009uw}, causal set theory \cite{Carlip:2015mra,Surya:2019ndm}, and others \cite{Carlip:2009km,Stojkovic:2013xcj,Niedermaier:2006wt}. Indeed, previous results hint that dimensional reduction might play a key role in regularizing gravity at short distances \cite{Stojkovic:2013xcj,Anchordoqui:2010er}. It is also worth noting that other frameworks have been introduced in the literature to model the physics of a variable number of dimensions \cite{Mansouri:1999kd,Saueressig:2021pzy,Calcagni:2017ymp,Calcagni:2021aap,Amelino-Camelia:2013tla,Amelino-Camelia:2013cfa,Brighenti:2022fvt}, including more general schemes such as multifractional geometry \cite{Calcagni:2009kc,Calcagni:2016xtk,Calcagni:2011sz}. These can be viewed as complementary to our approach.

Relying solely on the basic postulates of quantum mechanics (QM), our formalism provides a versatile and transparent way to investigate the emergence of physical dimensions. We anticipate further generalizations of the framework to relativistic quantum fields featuring various types of interactions across different dimensions. Furthermore, we envision possible applications across diverse fields, from condensed matter systems to cosmology and quantum gravity, where the dimension itself is treated within the QD framework. 

\emph{Dimension as a Quantum Variable---}A quantum system is characterized by a Hilbert space together with a set of operators acting on it. Suppose that for each number of spatial dimensions $d$ in the range $N_1\leq d\leq N_2$, with $N_1,N_2\in \mathbb{N}_0$ and $N_1\leq N_2$ we can define a quantum system with Hilbert space $\mathcal{H}^{(d)}$. This Hilbert space is spanned by the states
\begin{equation}\label{hilbertspacedconst}
    \mathcal{H}^{(d)}=\text{Span}\{\ket{i^{(d)}}|\  i^{(d)}\in I^{(d)}\},
\end{equation}
where $i^{(d)}\in I^{(d)}$ labels elements of an arbitrary basis of $\mathcal{H}^{(d)}$. We denote the scalar operators acting on the Hilbert space $\mathcal{H}^{(d)}$ by $\hat{O}^{(d)}$.

\begin{figure}
    \centering
    \includegraphics[width=\linewidth]{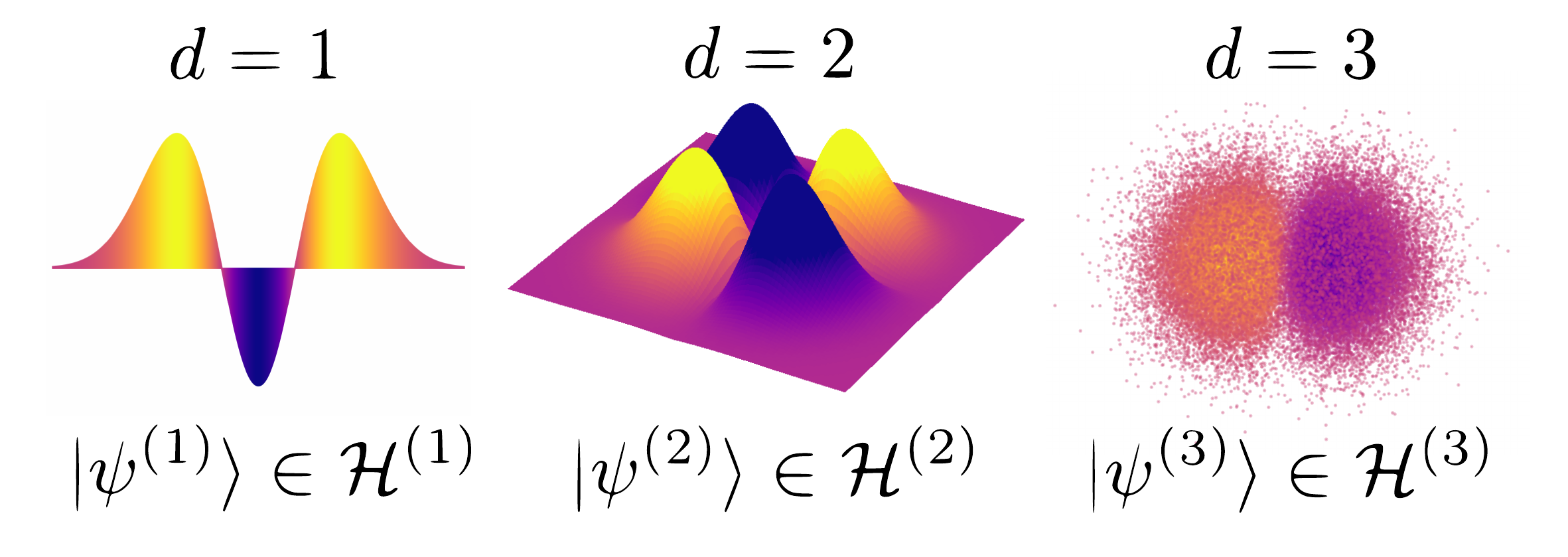}
    \caption{Graphical representation of a QD Hilbert space construction for a generalized harmonic oscillator. The three states $\ket{\psi^{(1)}}, \ket{\psi^{(2)}}, \ket{\psi^{(3)}}$ belong to $\mathcal{H}^{(1)}$, $\mathcal{H}^{(2)}$, and $\mathcal{H}^{(3)}$, respectively. The Hilbert space of the QD harmonic oscillator is then constructed by combining $\mathcal{H}^{(d)}$, $d=1,2,3$ via the direct sum $\bigoplus$. A generic state of a QD harmonic oscillator is a linear combination of states $\ket{\psi^{(d)}} \in \mathcal{H}^{(d)}$.}
    \label{fig:1} 
\end{figure}

A generic state of a QM system with a variable number of dimensions can be conceptualized as a superposition of states belonging to different $\mathcal{H}^{(d)}$ (see Fig.~\ref{fig:1}). Under this assumption, the Hilbert space of a QD system is taken to be
\begin{equation}\label{hilbertspace}
\mathcal{H}=\bigoplus_{d=N_1}^{N_2}\mathcal{H}^{(d)},
\end{equation}
where $\oplus$ denotes the direct sum. A generic state $\ket{\psi}\in\mathcal{H}$ can then be written as a superposition
\begin{equation}\label{genstat}
    \ket{\psi}=\sum_{d=N_1}^{N_2}\sum_{i^{(d)}\in I^{(d)}}c^{(d)}_{i^{(d)}}\ket{d;i^{(d)}},
\end{equation}
where ${c^{(d)}_{i^{(d)}}}$ are complex numbers and the additional index $d$ has been added to the kets in \eqref{hilbertspacedconst} to label the number of spatial dimensions in which the state resides. The coefficients ${c^{(d)}_{i^{(d)}}}$ satisfy the normalization condition $\braket{\psi|\psi}=1$, but are otherwise arbitrary. It follows that the QD Hilbert space $\mathcal{H}$ is a generalization of $\mathcal{H}^{(d)}$: the ordinary $d$-dimensional system is recovered in the limit $N_1=N_2=d$.

In analogy with \eqref{hilbertspace}, we can define generalized observables acting on the full Hilbert space $\mathcal{H}$ via the direct sum
\begin{equation}\label{genoperator}
\hat{O}=\bigoplus_{d=N_1}^{N_2}\hat{O}^{(d)}=
\begin{bmatrix}
\hat{O}^{(N_1)} & \cdots & 0\\
\vdots & \ddots & \vdots\\
0 & \cdots & \hat{O}^{(N_2)}\\
\end{bmatrix},
\end{equation}
where $\hat{O}^{(d)}$ are the ordinary $d$-dimensional operators. The action of the generalized operator \eqref{genoperator} on a state \eqref{genstat} is given by
\begin{equation}\label{acgen}
\bra{k;i^{(k)}}\hat{O}\ket{d;j^{(d)}}=
\begin{cases}
\bra{i^{(k)}}\hat{O}^{(d)}\ket{j^{(d)}} & \text{if} \ k=d, \\
0 & \text{if} \ k\neq d.
\end{cases}
\end{equation}
In addition to generalized observables \eqref{genoperator}, all QD systems are naturally endowed with a new ``dimension'' operator, an observable corresponding to the number of spatial dimensions:
\begin{equation}\label{dimop}
\hat{D}=\bigoplus_{d=N_1}^{N_2}d\hat{I}^{(d)}=\begin{bmatrix}
N_1\hat{I}^{(N_1)} & \cdots & 0\\
\vdots & \ddots & \vdots\\
0 & \cdots & N_2\hat{I}^{(N_2)}\\
\end{bmatrix},
\end{equation}
where $\hat{I}^{(d)}$ is the identity operator acting on $\mathcal{H}^{(d)}$. In the limit $N_1=N_2=d$, the generalized Hilbert space $\mathcal{H}\equiv\mathcal{H}^{(d)}$, and the dimension operator becomes trivial, i.e., proportional to the identity operator $\hat{I}^{(d)}$.

As in an ordinary system with a fixed number of dimensions, the time evolution is driven by the Hamiltonian. The most general QD Hamiltonian can be split into
\begin{equation}\label{totham1}
    \hat{H}=\hat{H}_0+\hat{H}_{int},
\end{equation}
where
\begin{equation}\label{totham2}
\hat{H}_0=\bigoplus_{d=N_1}^{N_2}\hat{H}^{(d)}
\end{equation}
is the generalization \eqref{genoperator} of the fixed-$d$ Hamiltonians $\hat{H}^{(d)}$ acting on $\mathcal{H}^{(d)}$, and $\hat{H}_{int}$ is a generic interaction term. In the case of a theory with $\hat{H}_{int}=0$, $\hat{D}$ commutes with the Hamiltonian:
\begin{equation}\label{HDcom}
\left[\hat{H}_0,\hat{D}\right]=\bigoplus_{d=N_1}^{N_2}d\left[\hat{H}^{(d)},\hat{I}^{(d)}\right]=0,
\end{equation}
and the dimensionality is a constant of motion. Conversely, if $\bra{d;i^{(d)}}\hat{H}_{int}\ket{k;j^{(k)}}\neq 0$ for $i^{(d)},j^{(k)}$ such that $d\neq k$, the Hamiltonian \eqref{totham1} may no longer commute with $\hat{D}$, allowing the dimensionality of the system to change dynamically.

\emph{QD Two-State System---}We now showcase our framework by constructing a simple two-state QD model. Consider a QD system that consists of two trivial subsystems in $d$ and $k$ spatial dimensions, each spanned by a single state \footnote{In the introduction, we assumed that the spatial dimension $d$ takes every integer value in the range $d = N_1,\dots,N_2$. In general, one is free to choose $d$ from any set of integer values—not necessarily a contiguous interval. For example, in the two-state QD system, the parameters $d$ and $k$ may be any non-negative integers and need not be adjacent.}:
\begin{equation}\label{trivsysthilb}
    \mathcal{H}^{(d)} = \mathrm{Span}\{\ket{0}\}, 
    \qquad
    \mathcal{H}^{(k)} = \mathrm{Span}\{\ket{0}\}.
\end{equation}
Their Hamiltonians are simply the $1\times 1$ matrices
\begin{equation}\label{trivsystham}
    \hat{H}^{(d)} = (E^{(d)}), 
    \qquad
    \hat{H}^{(k)} = (E^{(k)}),
\end{equation}
and the QD Hilbert space is, according to \eqref{hilbertspace}:
\begin{equation}
    \mathcal{H}
    = \mathcal{H}^{(d)} \oplus \mathcal{H}^{(k)}
    = \mathrm{Span}\{\ket{d;0}, \ket{k;0}\}.
\end{equation}
The full Hamiltonian consists of the $\hat{H}_0$ term:
\begin{align}\label{2stateham}
    \hat{H}_0 
    &= \hat{H}^{(d)} \oplus \hat{H}^{(k)}= 
    \begin{pmatrix}
        E^{(d)} & 0 \\
        0 & E^{(k)}
    \end{pmatrix},
\end{align}
and a generic interaction term, such that
\begin{align}\label{2stateham2}
    \hat{H}= \hat{H}_0+
    \begin{pmatrix}
        0 & \alpha \\
        \alpha & 0
    \end{pmatrix}=
    \begin{pmatrix}
        E^{(d)} & \alpha\\
        \alpha & E^{(k)}
    \end{pmatrix},
\end{align}
where the constant $\alpha$ introduces non-diagonal terms corresponding to $\hat{H}_{int}$. Without loss of generality, we can assume $\alpha$ to be real and non-negative, since any non-zero phase can be eliminated by a phase rotation of the basis states. 

The dimension operator in the basis $\{\ket{d;0},\ket{k;0}\}$ reads:
\begin{equation}
    \hat{D}=\begin{pmatrix}
d & 0 \\
0 & k
\end{pmatrix}.
\end{equation}
The two-state system is exactly solvable, and the Hamiltonian can be diagonalized explicitly, which results in the following eigenvalues:
\begin{equation}\label{DiagHam}
    E_{\pm} = \frac{E^{(d)}+E^{(k)}}{2} \pm \Omega, 
    \quad 
    \Omega = \sqrt{\left( \frac{E^{(d)}-E^{(k)}}{2} \right)^2 + \alpha^{2}}.
\end{equation}
The corresponding normalized eigenvectors are:
\begin{align}
    \ket{+} &= 
        \cos\theta \, \ket{d;0} 
        + \sin\theta \, \ket{k;0}, \\
    |-\rangle &= 
        -\,\sin\theta \, \ket{d;0} 
        +\cos\theta \, \ket{k;0},
\end{align}
with the mixing parameter $\theta$ given by \cite{Cohen-Tannoudji:101367}
\begin{equation}
    \tan(2\theta) = \frac{2\alpha}{E^{(d)}-E^{(k)}}.
\end{equation}

\begin{figure}[t!]
    \centering
    \includegraphics[width=\linewidth]{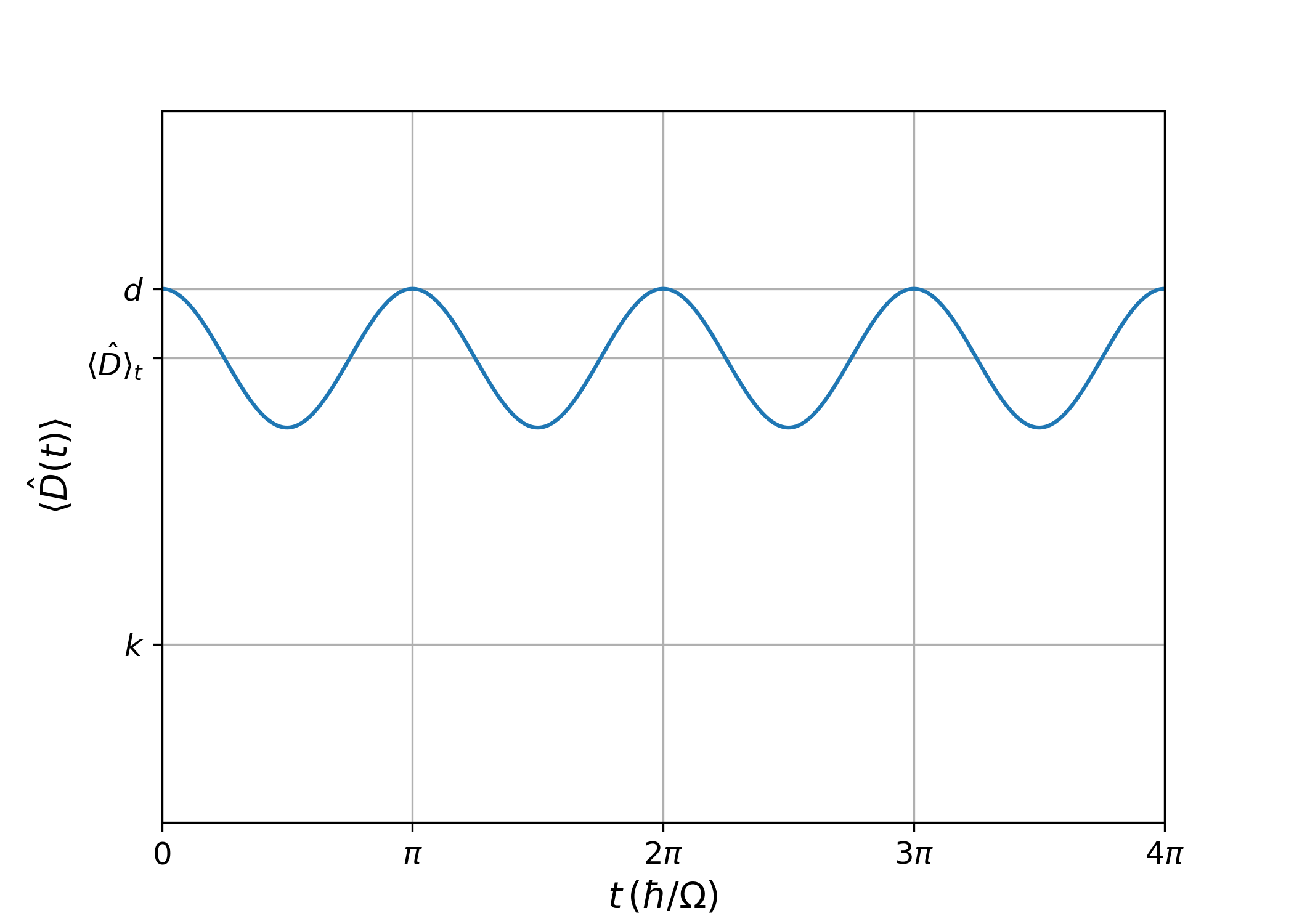}
    \caption{Oscillations in the dimension of a QD two-state system induced by the interaction term $\alpha$. The amplitude of the oscillations is $(k-d)\frac{\alpha^2}{2\Omega^2}$, and the frequency is $2\Omega/\hbar$.}
    \label{fig:2}
\end{figure}

Now, suppose that at the time $t=0$ the interacting system is measured to be in $d$ spatial dimensions, i.e. in the state $\ket{\psi(0)}=\ket{d;0}$. In order to see how the dimension of the system evolves in time, we expand the initial state in the interacting basis $\ket{+}, \ket{-}$:
\begin{equation}
    \ket{\psi(0)}=\ket{d;0} = \cos\theta \, \ket{+} 
                 - \sin\theta \, \ket{-},
\end{equation}
hence the state evolves in time according to
\begin{equation}
    |\psi(t)\rangle 
    = \cos\theta \, e^{-\frac{i}{\hbar}E_{+} t} \ket{+} 
      - \sin\theta \, e^{-\frac{i}{\hbar}E_{-} t} \ket{-} .
\end{equation}
The expectation value of the dimension operator $\hat{D}$ then reads:
\begin{equation}
    \bra{\psi(t)}\hat{D}\ket{\psi(t)}
    = d + (k-d)\frac{\alpha^{2}}{2\Omega^{2}}
        \left[1 - \cos\left(\frac{2\Omega}{\hbar} t\right)\right].
\end{equation}
This result shows explicitly that the perturbation $\alpha$ induces coherent oscillations in the observable number of dimensions at frequency $2\Omega/\hbar$ with an amplitude proportional to $\alpha^{2}/\Omega^{2}$. For small $\alpha$, the dimension of the system oscillates with a small amplitude $\propto\alpha^2$ around the time-averaged value
\begin{equation}
    \braket{\hat{D}}_t=d+(k-d)\frac{\alpha^2}{2\Omega^2}.
\end{equation}
The time dependence of the dimension of the system is plotted in Fig.~\ref{fig:2}.

Suppose that the two-state QD system is brought to thermal equilibrium with temperature $T=\frac{1}{k_B \beta}$ (for simplicity, we will assume the system to be non-degenerate, i.e. $E_{+}\neq E_{-}$). The system is therefore described by the canonical distribution:
\begin{equation}
Z(\beta)=\mathrm{Tr}\left(e^{-\beta \hat{H}}\right)
= e^{-\beta E_{+}}+e^{-\beta E_{-}}.
\end{equation}
The thermal expectation values of the Hamiltonian and the dimension operators are:
\begin{equation}
\begin{aligned}
&\braket{\hat{H}}=\frac{\mathrm{Tr}(e^{-\beta \hat{H}}\hat{H})}{Z(\beta)}=\frac{E_{+}e^{-\beta E_{+}}+E_{-}e^{-\beta E_{-}}}{e^{-\beta E_{+}}+e^{-\beta E_{-}}},\\ 
&\braket{\hat{D}}=\frac{\mathrm{Tr}(e^{-\beta \hat{H}}\hat{D})}{Z(\beta)}=\frac{e^{-\beta E_{+}}(d\cos^2(\theta)+k\sin^2(\theta))}{e^{-\beta E_{+}}+e^{-\beta E_{-}}}\\
&+\frac{e^{-\beta E_{-}}(d\sin^2(\theta)+k\cos^2(\theta))}{e^{-\beta E_{+}}+e^{-\beta E_{-}}}.
\end{aligned}
\end{equation}
The effective dimension $\braket{\hat{D}}$ depends on the inverse temperature $\beta$ (see Fig.~\ref{fig:3}). The relation with the thermal average of the Hamiltonian reads:

\begin{figure}[t!]
    \centering
    \includegraphics[width=\linewidth]{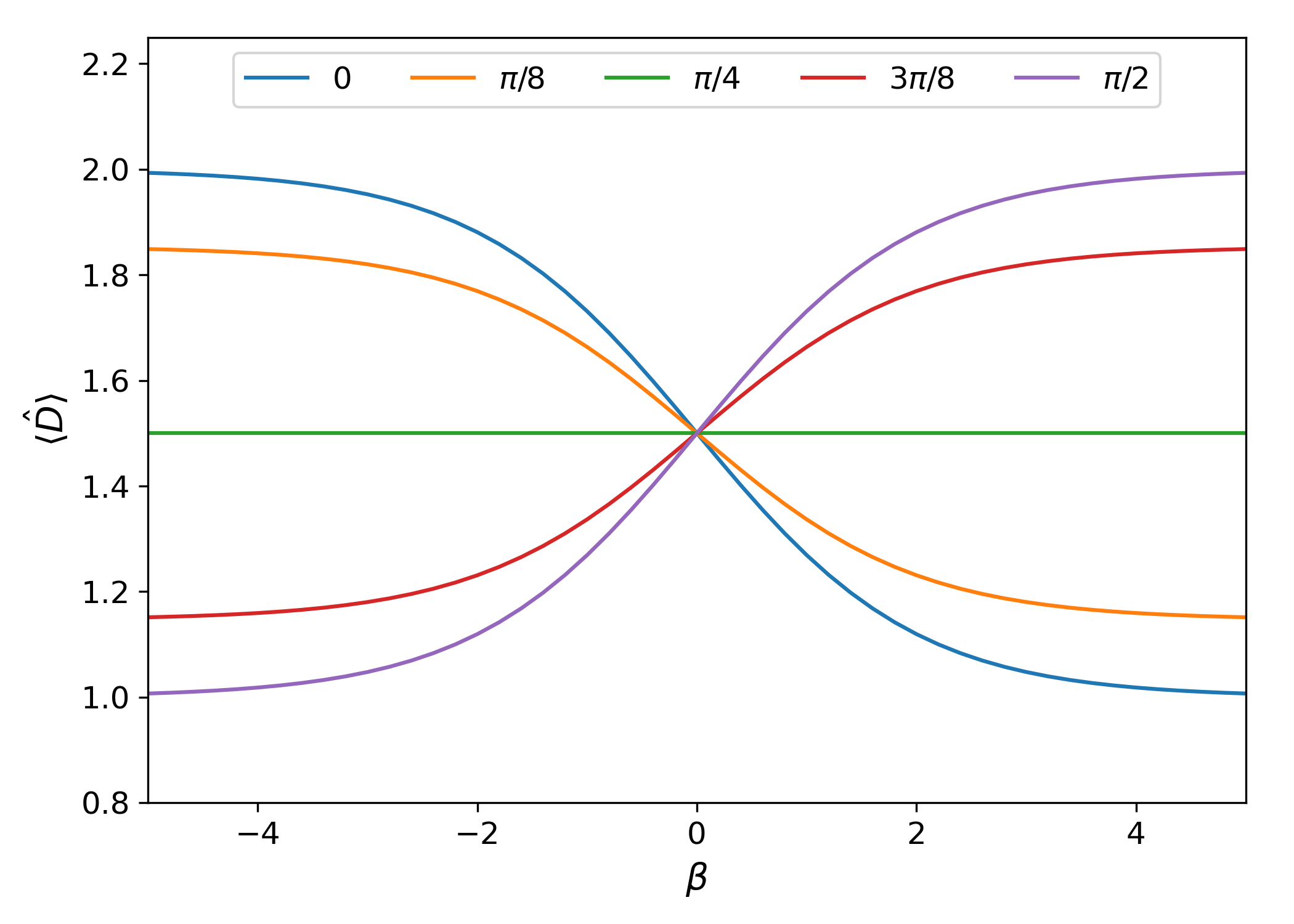}
    \caption{The plot of the thermal expectation value of the dimension operator $\hat{D}$ as a function of $\beta$ for five different values of the mixing parameter $\theta$. The plot was generated for a two-state QD system with energies $E_+=2, E_-=1$ and dimensions $d=2, k=1$. Depending on the value of the coupling constant, $\braket{\hat{D}}$ either increases or decreases with $\beta$.}
    \label{fig:3}
\end{figure}

\begin{equation}\label{DfuncE}
\begin{aligned}
&\braket{\hat{D}}=\frac{(d-k)\cos(2\theta)}{E_{+}-E_{-}}\braket{\hat{H}}\\
&+\frac{(d\sin^2(\theta)+k\cos^2(\theta))E_{+}-(d\cos^2(\theta)+k\sin^2(\theta))E_{-}}{E_+-E_-}.
\end{aligned}
\end{equation}

The dimension $\braket{\hat{D}}$ of the system increases or decreases linearly with $\braket{\hat{H}}$, depending on the energy levels $E_+$ and $E_-$. This example demonstrates one possible mechanism by which a QD system may experience scale-dependent variations in the number of spatial dimensions. It is also worth noting that the linear relation \eqref{DfuncE} is due to the simplicity of a two-state system, and in general $\braket{\hat{D}}$ can be a more complicated function of $\braket{\hat{H}}$. An alternative example that exhibits a richer relation is provided in the End Matter.

Even though we started with an integer-valued spectrum for $\hat{D}$, the effective dimension $\braket{\hat{D}}$ need not be an integer and can vary continuously with the energy scale. It should also be emphasized that $\braket{\hat{D}}$ is only one of the possible dimensional estimators. In principle, one could define dimensionality based on other physical quantities \cite{Carlip:2017eud}. A notable advantage of our framework is access to various dimensional estimators from first principles.

\emph{Enhanced Symmetries---}We now turn to the symmetries supported by the enlarged Hilbert space of QD systems. A direct generalization of the symmetries already present in the fixed-$d$ case is given by prescription \eqref{genoperator}. Suppose that each Hamiltonian $\hat{H}^{(d)}$ is invariant under a unitary symmetry acting on $\mathcal{H}^{(d)}$ via an operator $\hat{U}^{(d)}$. Then, the QD system is invariant under a combined symmetry represented by the operator:
\begin{equation}\label{symoper}
    \hat{U}=\bigoplus_{d=N_1}^{N_2}\hat{U}^{(d)}.
\end{equation}
This can be considered a form of subsystem symmetry \cite{Moudgalya:2021ixk,Moudgalya:2022gtj}, where each $\hat{U}^{(d)}$ acts in the usual way on the $\mathcal{H}^{(d)}$ subspace of $\mathcal{H}$ and trivially on the remaining states. The system is therefore invariant under the symmetry group
\begin{equation}\label{gengroup}
    G=\prod_{d=N_1}^{N_2}G^{(d)},
\end{equation}
where $G^{(d)}$ are the symmetry groups acting through $\hat{U}^{(d)}$ on the fixed-$d$ subspaces $\mathcal{H}^{(d)}$. Notably, QD systems can also exhibit additional degeneracies not accounted for by a symmetry of the form \eqref{gengroup}. This occurs when the spectra of Hamiltonians in $d$ and $k$ dimensions have common eigenvalues for $d\neq k$. This is precisely the case of the two-state QD system discussed in the previous section. For any two values of $E^{(d)}$ and $E^{(k)}$, the system supports a $\mathrm{U}(1)\times \mathrm{U}(1)$ symmetry corresponding to a phase shift of the basis vectors $\ket{d;0}$ and $\ket{k;0}$. This symmetry is an example of a symmetry \eqref{gengroup}, inherited from the global phase symmetries of $\hat{H}^{(d)}$ and $\hat{H}^{(k)}$.

Now, suppose that $\alpha=0$ and the two energy eigenvalues are equal, i.e. $E^{(d)}=E^{(k)}$. This creates a $2$-fold degeneracy in $\hat{H}_0$, with states of different dimensionalities belonging to the same energy level. It can be checked that in this case, a unitary transformation $\hat{U}$ acting on the basis states via:
\begin{equation}\label{transfenhsym}
\begin{aligned}
&\ket{d;0}\rightarrow U_{11}\ket{d;0}+U_{12}\ket{k;0},\\
&\ket{k;0}\rightarrow U_{21}\ket{d;0}+U_{22}\ket{k;0},
\end{aligned}
\end{equation}
leaves the Hamiltonian invariant
\begin{equation}
    \hat{U}\hat{H}_0\hat{U}^{-1}=\hat{H}_0,
\end{equation}
and therefore the full symmetry is enhanced to $\mathrm{U}(2)$ \cite{Cariglia:2014ysa,Calixto:2019qwm}. In the limit $N_1=N_2=d$ (i.e. $d=k$), when we recover an ordinary QM system, this additional symmetry reduces to a $\mathrm{U}(1)$ phase transformation on the eigenstates of $\mathcal{H}$.

Note that the symmetry operator represented by $U$ mixes states with different dimensionality, in contrast with the simple generalized symmetry \eqref{gengroup}, and therefore corresponds to a new symmetry not present in regular QM systems. Symmetries of this form provide a map between the lower- and higher-dimensional states of the Hilbert space. For this reason, they could potentially be useful for understanding higher-dimensional systems in terms of their lower-dimensional counterparts.

However, the presence and structure of this additional symmetry typically depend on how the system is defined. In ordinary QM, shifting the vacuum energy by a constant is not directly measurable. However, when several such ordinary QM systems are assembled together into a QD system according to the rules \eqref{hilbertspace} and \eqref{totham1}, individual constant shifts in the fixed-$d$ Hamiltonians can result in observable effects, such as an altered symmetry structure. The situation is analogous to a case in which an ordinary QM system is constructed from two subsystems $A$ and $B$. If the systems are interacting, one cannot offset the vacuum energy of systems $A$ or $B$ individually without affecting the physics. The system is invariant only under the vacuum shift of the combined system, i.e., when the energy levels of $A$ and $B$ shift by the same amount.

\emph{Discussion---}In this Letter, we have proposed and explored a quantum framework in which the dimensionality is promoted to a quantum operator $\hat{D}$. To illustrate our idea, we examined a two-state QD model with a variable number of dimensions. We solved the system analytically and computed the partition function, which encodes how the effective dimensionality varies with temperature. This provides a natural example of dimensionality being dependent on the energy scale of the system. Moreover, we have shown that QD systems can exhibit enhanced symmetries which are sensitive to the vacuum energies of the fixed-$d$ subsystems. The formalism introduced in this work relies only on a few basic assumptions, making it sufficiently versatile to explore a wide class of QD systems.

In the future, implementing the framework in gravitational systems could provide a simple way to model dimensional reduction and examine its physical consequences. More broadly, the formalism is of relevance in quantum field theory, where the Hilbert space $\mathcal{H}$ could be constructed from the appropriate QD multi-particle spaces. Building on the arguments discussed previously, we speculate that such QD systems could exhibit improved renormalization properties. Finally, our framework applies to condensed matter systems whose effective dimensionality depends on the energy scale. Further exploration of interactions within our framework could potentially lead to entirely new quantum systems with intriguing physical properties.\break

\emph{Acknowledgements---}We would like to thank Vania Vellucci, Manuel Del Piano, and Zeyu Ma for helpful discussions. This work was partially supported by the Carlsberg Foundation under the grant CF22-0922.

\nocite{*}

\bibliography{references}

\clearpage

\onecolumngrid   
\section*{End Matter}
\twocolumngrid 

\newpage
\emph{QD Harmonic Oscillator---} Here, we show how the framework generalizes to physical systems with a richer energy spectrum: the harmonic oscillator. The Hamiltonian of an ordinary harmonic oscillator in $d$ spatial dimensions can be written as
\begin{equation}\label{hamregosc}
    \hat{H}^{(d)}=\sum_{i=1}^{d}\left((\hat{a}_i^{(d)})^{\dagger}\hat{a}_i^{(d)}
    +\frac{1}{2}\hat{I}^{(d)}\right),
\end{equation}
where $(\hat{a}_i^{(d)})^{\dagger}$ and $\hat{a}_i^{(d)}$ are the $d$-dimensional creation and annihilation operators, respectively \cite{Shankar1994}. For simplicity, we adopt units in which $\hbar\omega=1$ and set the energy such that the ground state of the system is at $E=d/2$. After diagonalizing, the eigenstates and eigenvalues of \eqref{hamregosc} are labeled by a set of $d$ integers $k_1^{(d)},\ldots,k_d^{(d)}\in\mathbb{N}_0$:
\begin{equation}\label{specreg}
\ket{k_1^{(d)},\ldots,k_d^{(d)}}, \quad E(k_1^{(d)},\ldots,k_d^{(d)})=\frac{d}{2}+
\sum_{i=1}^{d}k_i^{(d)}.
\end{equation}

Consider a QD generalization of \eqref{hamregosc} in the case where the Hamiltonian \eqref{totham1} does not have an interacting term $\hat{H}_{int}$. The generalized Hamiltonian is therefore constructed solely using \eqref{totham2}:
\begin{equation}\label{genham}
\hat{H}_0=\bigoplus_{d=N_1}^{N_2}\sum_{i=1}^{d}\left((\hat{a}_i^{(d)})^{\dagger}\hat{a}_i^{(d)}+\frac{1}{2}\hat{I}^{(d)}\right).
\end{equation}
The eigenstates of the Hamiltonian are again \eqref{specreg}, but now $d$ ranges over all possible values:
\begin{equation}\label{genstates}
\ket{d;k_1^{(d)},\ldots,k_d^{(d)}}, \quad N_1\leq d\leq N_2, \quad k_1^{(d)},\ldots,k_d^{(d)}\in\mathbb{N}_0.
\end{equation}
The spectrum of $\hat{H}_0$ is then given by:
\begin{equation}
E(k_1^{(d)},\ldots,k_d^{(d)})=\frac{d}{2}+\sum_{i=1}^{d}k_i^{(d)}.
\end{equation}

In thermal equilibrium at the inverse temperature $\beta$, the system is described by the partition function
\begin{equation}
Z(\beta)=\sum_{d=N_1}^{N_2}\left(Z^{(1)}(\beta)\right)^d, \quad Z^{(1)}(\beta)=\frac{e^{-\beta/2}}{1-e^{-\beta}}.
\end{equation}
where $Z^{(1)}(\beta)$ is the partition function of an ordinary $1$-dimensional harmonic oscillator. Using the canonical distribution, the expectation values of the generalized Hamiltonian \eqref{genham} and the dimension operator can be written as:
\begin{figure}[H]
    \centering
    \includegraphics[width=\linewidth]{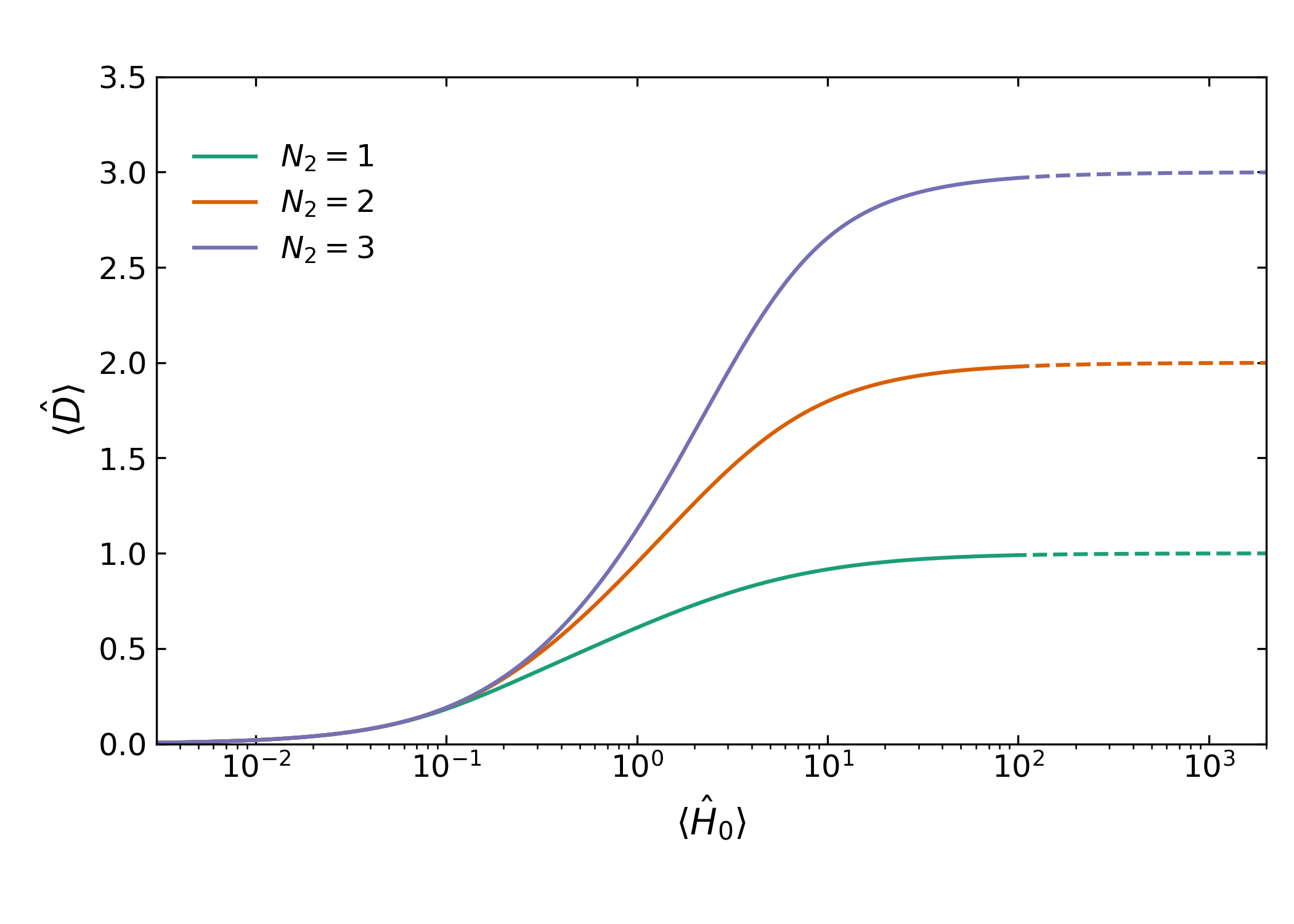}
    \caption{The number of spatial dimensions for three QD harmonic oscillators with $N_1=0$ plotted as a function of energy for three different values of $N_2$. The expectation value $\braket{\hat{D}}$ increases with the energy scale, starting from $\braket{\hat{D}}=N_1=0$ in the limit $\braket{\hat{H}_0}=0$ (dotted line) and approaching $\braket{\hat{D}}=N_2$ at high energies (dashed line).}
    \label{fig:4}
\end{figure}
\begin{equation}
\braket{\hat{H}_0}=-\frac{1}{Z}\partial_{\beta}Z, \quad \braket{\hat{D}}=\frac{\sum_{d=N_1}^{N_2}d(Z^{(1)})^d}{\sum_{d=N_1}^{N_2}\left(Z^{(1)}\right)^d}.
\end{equation} 

The plot of $\braket{\hat{D}}$ as a function of $\braket{\hat{H}_0}$ is shown in Fig.~\ref{fig:4}. As opposed to the simple two-state QD system, $\braket{\hat{D}}$ depends nonlinearly on the thermal energy $\braket{\hat{H}_0}$.

The QD harmonic oscillator also exhibits additional degeneracies that are not explained by the symmetry \eqref{gengroup}. Consider a QD harmonic oscillator with $N_1=0$ and $N_2=3$, which corresponds to the purple line in Fig.~\ref{fig:4}. The third excited state $E=3/2$ is $2$-fold degenerate:
\begin{equation}\label{groundstate}
E=\frac{3}{2}: \ket{1;1}, \ \ket{3;0,0,0},
\end{equation}
with states of different dimensionalities belonging to the same energy level. It is easy to verify that the transformation
\begin{equation}\label{transfenhsym2}
\begin{aligned}
&\ket{1;1}\rightarrow U_{11}\ket{1;1}+U_{12}\ket{3;0,0,0},\\
&\ket{3;0,0,0}\rightarrow U_{21}\ket{1;1}+U_{22}\ket{3;0,0,0},
\end{aligned}
\end{equation}
where the complex numbers $U_{ij}$ form a $2\times2$ unitary matrix $U$ that leaves the Hamiltonian $\hat{H}_0$ invariant. More generally, any unitary mixing within a given degeneracy of the Hamiltonian \eqref{genham} will leave it invariant. The QD harmonic oscillator provides another example of a system with an energy-dependent effective dimension. Furthermore, it illustrates how the QD framework leads to physically rich dynamics in which new symmetries intertwine states across different dimensions.

\end{document}